\magnification \magstep1
\input amssym.def
\input amssym.tex
\bigskip
\centerline{\bf Vaidya Space-Time in Black-Hole Evaporation}
\bigskip
\centerline{ A.N.St.J.Farley and P.D.D'Eath}
\bigskip
\smallskip
\centerline{Department of Applied Mathematics and Theoretical Physics,
Centre for Mathematical Sciences,} 
\smallskip
\centerline{University of Cambridge, Wilberforce Road, Cambridge CB3 0WA,
United Kingdom}
\bigskip
\centerline{Abstract}
\smallskip
This paper continues earlier work on the quantum evaporation of black
holes.  This work has been concerned with the calculation and 
understanding of quantum amplitudes for final data perturbed slightly 
away from spherical symmetry on a space-like hypersurface $\Sigma_F$ 
at a late Lorentzian time $T{\,}$.  For initial data, we take, for 
simplicity, spherically-symmetric asymptotically-flat data for Einstein 
gravity with a massless scalar field on an initial surface $\Sigma_I$ 
at time $t=0{\,}$.  Together, such boundary data give a quantum
analogue of classical Einstein/scalar gravitational collapse to a
black hole, perhaps starting from a diffuse, early-time configuration.  
Quantum amplitudes are calculated following Feynman's approach, by 
first rotating:
$T\rightarrow{\mid}T{\mid}\exp(-i\theta)$ into the complex, where 
$0<\theta\leq\pi/2{\,}$, then solving the corresponding complex 
{\it classical} boundary-value problem, which is expected to be 
well-posed provided ${\theta}>0{\,}$, and computing its classical 
Lorentzian action $S_{\rm class}$ and corresponding semi-classical 
quantum amplitude, proportional to $\exp(iS_{\rm class})$.  For a
locally-supersymmetric Lagrangian, describing supergravity coupled to
supermatter, any loop corrections will be negligible, provided that the
frequencies involved in the boundary data are well below the Planck
scale.  Finally, the Lorentzian amplitude is recovered by taking the 
limit $\theta\rightarrow 0_{+}$ of the semi-classical amplitude.  
In the black-hole case, by studying the linearised spin-0 or spin-2 
{\it classical} solutions in the above (slightly complexified) case, 
for the corresponding classical boundary-value problem with the given
perturbative data on $\Sigma_F{\,}$, one can compute an effective 
energy-momentum tensor $<T^{\mu\nu}>_{EFF}{\,}$, which has been 
averaged over several wavelengths of the radiation, and which
describes the averaged extra energy-momentum contribution in the 
Einstein field equations, due to the perturbations.  In general, this 
averaged extra contribution will be spherically symmetric, being of
the form of a null fluid, describing the radiation (of quantum origin) 
streaming radially outwards.  The corresponding space-time metric, in 
this region containing radially outgoing radiation, is of the Vaidya
form.  This, in turn, justifies the treatment of the adiabatic radial 
mode equations, for spins $s=0$ and $s=2{\,}$, which is used elsewhere
in this work.\par 
\medskip
\noindent
{\bf 1. Introduction}
\medskip
\indent
This paper is concerned with the problem of finding approximate
classical Lorentzian (or slightly complexified) solutions of the
coupled Einstein gravity/massless-scalar field equations, to describe
the region of space-time containing outgoing radiation (both spin-0
and spin-2) in a very large number of modes, generated by
quantum-mechanical evaporation, as a result of nearly-spherical
gravitational collapse to a black hole.  The space-time metric
$g_{\mu\nu}$ and scalar field $\phi$ are split into a 'background'
spherically-symmetric part $(\gamma_{\mu\nu}{\,},\Phi)$, plus
perturbations ${\,}(h^{(1)}_{\mu\nu}{\,},\phi^{(1)}){\,}$, etc., 
which are typically non-spherical.  The energy-momentum tensor formed 
from $\Phi$ provides the 'matter source' for an exactly spherical 
collapse to a black hole.\par
\smallskip
\indent
In recent papers [1-5], the quantum amplitude for a given 
perturbative configuration (say of the scalar field $\phi^{(1)}$) 
on a final hypersurface $\Sigma_F$ at a very late time $T{\,}$, was 
found by rotating $T$ slightly into the complex: 
$T\rightarrow{\mid}T{\mid}\exp(-i\theta)$, for
${\,}0<\theta\leq\pi/2{\,}$; then calculating the (complex-valued)
Lorentzian classical action $S_{\rm class}$ for the corresponding 
{\it classical} boundary-value problem, which is expected to be
well-defined; then computing the resulting semi-classical amplitude,
proportional to ${\,}\exp(iS_{\rm class})$;  then finally obtaining the
amplitude for real Lorentzian $T$ by taking the limit of 
${\,}\exp(iS_{\rm class})$ as ${\,}\theta\rightarrow{\,}0_{+}{\,}$.
Typically, the perturbative scalar-field configuration $\phi^{(1)}$
given on the late-time surface $\Sigma_F$ will involve an enormous
number of modes, both angular and radial, but with a minute
coefficient for each such mode.  That is, the given $\phi^{(1)}$ may
contain extremely detailed angular structure, and also be spread over
a considerable radius from the centre of spherical symmetry of the
background $(\gamma_{\mu\nu}{\,},\Phi)$, again with detailed radial
structure.  Now consider the corresponding classical Dirichlet 
boundary-value problem above, in which one takes 
$\phi^{(1)}=\phi^{(1)}{\mid}_{\Sigma_F}$ as given on $\Sigma_F{\,}$, 
but chooses ${\,}\phi^{(1)}{\mid}_{\Sigma_I}=0$ (for simplicity), 
together with the complex time-interval-at-infinity 
${\,}T={\mid}T{\mid}\exp(-i\theta){\,}$ for ${\,}0<\theta\leq\pi/2{\,}$.  
The solution for $\phi^{(1)}$ will gradually decay towards zero, as
one moves from the final surface $\Sigma_F$ to earlier times; the rate 
of this exponential decay will be extremely slow when ${\,}\theta$ is 
close to zero.  In this case, one will find that, at all times $t$
with ${\,}0<t<T$ (that is, between $\Sigma_I$ and $\Sigma_F$), the 
classical solution will continue to have complicated angular and 
radial structure, much as does its boundary value 
$\phi^{(1)}{\mid}_{\Sigma_F}{\,}$.\par
\smallskip
\indent
As a result, one must study Lorentzian (or complexified Lorentzian)
classical solutions for the linearised metric and scalar perturbations
$(h^{(1)}_{\mu\nu}{\,},\phi^{(1)})$, which contain classical spin-0
and spin-2 radiation outgoing from the 'gravitational collapse', with
detailed structure over (typically) an enormous radial extent.  The
cumulative effective energy-momentum tensor, formed quadratically from
derivatives of these first-order perturbations, and then averaged over
several wavelengths of the radiation, so as to produce a smooth
averaged $<T^{\mu\nu}>_{EFF}{\,}$, is expected to be nearly 
spherically-symmetric, and indeed to have the form appropriate to a 
radially-outgoing null fluid [6].  This viewpoint simplifies
enormously the description of the 'effective energy-momentum source' 
due to the wave-like perturbations, which then feeds back into the
spherically-symmetric background solution 
$(\gamma_{\mu\nu}{\,},\Phi)$, albeit over a suitably long time-scale.  
In this description, the effective energy-momentum contribution of the 
emitted radiation can be reduced to just one spherically-symmetric 
'density of radiation' function of retarded time, instead of an 
infinite number of multipole or mode coefficients for the final boundary data
${\,}\phi^{(1)}{\mid}_{\Sigma_F}{\,}$.\par
\smallskip
\indent
The space-metric metric resulting from such a null-fluid effective
$T_{\mu\nu}$ is precisely of the Vaidya type [7].  This resembles the
Schwarzschild geometry, except that the role of the Schwarzschild mass
$M$ is taken by a mass function $m(t,r)$, which varies extremely
slowly with respect both to $t$ and to $r$ in the space-time region
containing outgoing radiation.  In this region, the slowly-varying
Vaidya metric provides a valid approximation.  Of course, one does not 
expect such a relatively simple analytic approximation to the metric 
and scalar field in the strong-field collapse region, where, in the
case of a real time-interval $T{\,}$, the classical Lorentzian 
black-hole solution is highly dynamical.\par
\smallskip
\indent
In Sec.2, we discuss the calculation and form of the averaged
energy-momentum tensor $< T^{\mu\nu}>_{EFF}{\,}$, assuming that both 
spin-0 perturbations $\phi^{(1)}$ and spin-2 (graviton) perturbations
$h^{(1)}_{\mu\nu}$ are present.  It is consistently assumed that the
time-scale associated with typical radiation frequencies is very much
less than the time-scale over which the background geometry changes.
Thus, the wave-like perturbations in the metric and in the scalar
field can be treated in a WKB approximation, leading to an expression
for $<T^{\mu\nu}>_{EFF}{\,}$.  One can then verify that this
${\,}<T^{\mu\nu}>_{EFF}$ generates an extremely slow evolution of the
resulting Vaidya metric.  The case of spin-1 (Maxwell) perturbations 
is also discussed.  The resulting Vaidya metric in the 
outgoing-radiation region is described in Sec.3, in different
coordinate systems adapted to different aspects of the radiating
system; this material has also been covered in part in [20].  
A brief Conclusion is included in Sec.4.\par
\medskip
\noindent
{\bf 2. High-frequency limit: fields and energy-momentum tensor}
\medskip
\indent
In Sec.3 of [3], we expanded out the Einstein field equations in
powers of $\epsilon{\,}$, given a perturbative expansion for the 
classical solution $(g_{\mu\nu}{\,},\phi)$ about a 
spherically-symmetric reference or 'background' solution 
$(\gamma_{\mu\nu}{\,},\Phi)$.  We write
$$g_{\mu\nu}(x,\epsilon){\;}{\,}
={\;}{\,}\gamma_{\mu\nu}(x){\;}
+{\,}\epsilon{\,}h^{(1)}_{\mu\nu}(x){\;}
+{\,}\epsilon^2{\,}h^{(2)}_{\mu\nu}(x){\;}+{\;}\ldots{\quad},
\eqno(2.1)$$
$$\phi(x,\epsilon){\;}{\,}
={\;}{\,}\Phi(\tau,r){\;}+{\,}\epsilon{\,}\phi^{(1)}(x){\;}
+{\,}{\epsilon}^{2}{\,}\phi^{(2)}(x){\;}+{\;}\ldots{\quad}.
\eqno(2.2)$$
\noindent
At lowest order $O({\epsilon}^{0})$, one has the background Einstein and
scalar field equations
$$\eqalignno{R^{(0)}_{\mu\nu}{\,}
-{\,}{1\over 2}{\,}R^{(0)}{\,}\gamma_{\mu\nu}{\;}{\,}&
={\;}{\,}8\pi{\,}T^{(0)}_{\mu\nu}{\quad}, &(2.3)\cr
\gamma^{\mu\nu}{\,}\Phi_{;\mu\nu}{\;}{\,}&
={\;}{\,}0{\quad}, &(2.4)\cr}$$
\noindent
Here, $R^{(0)}_{\mu\nu}$ denotes the Ricci tensor and $R^{(0)}$ denotes
the Ricci scalar of the background geometry $\gamma_{\mu\nu}{\,}$. 
Covariant differentiation in the background is denoted by a semi-colon 
$(\;\;)_{;\alpha}$ or (below) by $\nabla_{\alpha}(\;\;)$.  The 
background energy-momentum tensor is denoted by
$$T^{(0)}_{\mu\nu}{\;}{\,}
={\;}{\,}\Phi,_{\mu}{\,}\Phi,_{\nu}{\,}
-{\,}{1\over 2}{\,}\gamma_{\mu\nu}{\,}
\bigl(\Phi,_{\alpha}\Phi,_{\beta}{\,}\gamma^{\alpha\beta}\bigr).
\eqno (2.5)$$
\indent
The linearised or $O(\epsilon^{1})$ part of the Einstein field
equations reads [9]
$$\eqalign{\bar{h}^{(1)~~~;\sigma}_{\mu\nu;\sigma}  
&-2{\,}\bar{h}^{(1)~~~;\sigma}_{\sigma(\mu ;\nu)} 
-2{\,}R^{(0)}_{\sigma\mu\nu\alpha}{\,}\bar{h}^{(1)\sigma\alpha} 
-2{\,}R^{(0)\alpha}{}_{(\mu}\bar{h}^{(1)}_{\nu)\alpha}\cr 
+{\,}&\gamma_{\mu\nu}{\,}
\bigl({\,}\bar{h}^{(1)~~;\alpha\beta}_{\alpha\beta}
-{\,}\bar{h}^{(1)\alpha\beta}{\;}R^{(0)}_{\alpha\beta}{\,}\bigr)
+{\,}\bar{h}^{(1)}_{\mu\nu}{\,}R^{(0)}{\;}{\,}
={\;}{\,}-16\pi{\,}T^{(1)}_{\mu\nu}{\quad}.\cr}\eqno(2.6)$$
\noindent
where indices on all quantities are raised and lowered using the
background metric $\gamma_{\mu\nu}{\,}$.  As usual [9], we define 
$$\bar{h}^{(1)}_{\mu\nu}{\;}{\,}
={\;}{\,}h^{(1)}_{\mu\nu}{\,}
-{\,}{1\over 2}{\,}\gamma_{\mu\nu}{\;}h^{(1)}{\quad},\eqno(2.7)$$
\noindent
where 
$$h^{(1)}{\;}{\,}={\;}{\,}h^{(1)\mu}_{\mu}{\quad}.\eqno(2.8)$$
\noindent
Here, $R^{(0)}_{\sigma\mu\nu\alpha}$ denotes the Riemann tensor of the
background geometry $\gamma_{\mu\nu}{\,}$, and $T^{(1)}_{\mu\nu}$
denotes the linearisation or $O(\epsilon^1)$ part of the
energy-momentum tensor $T_{\mu\nu}(x,\epsilon)$, given explicitly in
Eq.(3.23) of [3].\par
\smallskip
\indent
The linearised Einstein equations (2.6) are most easily studied 
in a 'linearised harmonic gauge' [9] in which, by an infinitesimal
coordinate transformation, one has arranged that
$${\bar h}^{(1);\alpha}_{\alpha\beta}{\;}{\,}={\;}{\,}0{\quad}.\eqno(2.9)$$ 
At very late Lorentzian times, the background Riemann curvature and
the background scalar field $\Phi$ will die off rapidly, whence the
linearised Einstein equations (2.6) simplify to
$${\bar h}^{(1)~~;\sigma}_{\mu\nu;\sigma}{\;}
-{\;}2{\;}{\bar h}^{(1)~;\sigma}_{\sigma(\mu ~~;\nu)}{\;} 
-{\;}2{\;}R^{(0)}_{\sigma\mu\nu\alpha}{\;}{\bar h}^{(1)\sigma\alpha}{\;}
+{\;}\gamma_{\mu\nu}{\;}{\bar h}^{(1);\alpha\beta}_{\alpha\beta}{\;}{\,}
={\;}{\,}0{\quad},\eqno(2.10)$$
\noindent
that is, the linearised vacuum field equations [9], subject also to
Eq.(2.9).\par
\smallskip
\indent
As described in [4], the rate of change with time of the
spherically-symmetric background geometry $\gamma_{\mu\nu}$ will be 
extremely small, during the long quasi-static period when the rate of 
emission of quantum radiation by the black hole hardly varies with
time.  Hence, most perturbation modes, for scalar (spin-0) or 
gravitational (spin-2) oscillations, will be 'adiabatic' or 
high-frequency.  Within the high-frequency approximation, in a 
space-time without background matter, one may additionally 
(without loss of generality) impose the traceless gauge condition [9]:
$${{\bar h}^{(1)\alpha}}_{~~~~\alpha}{\;}{\,}={\;}{\,}0{\quad}.\eqno(2.11)$$
In this case, the linearised Einstein field equations (2.6), subject to
the transverse-traceless $(TT)$ gauge conditions (2.9,11), reduce 
further [9] to:
$$h^{(1)~~;\sigma}_{\mu\nu;\sigma}
-{\,}2{\,}R^{(0)}_{\sigma\mu\nu\alpha}{\;}h^{(1)\sigma\alpha}{\;}{\,} 
={\;}{\,}0{\quad}.\eqno(2.12)$$
\smallskip
\indent
At $O(\epsilon^{2})$, the gravitational field equations give the
second-order contribution $G^{(2)}_{\mu\nu}$ to the Einstein tensor
$$G_{\mu\nu}{\;}{\,}
={\;}{\,}R_{\mu\nu}{\,}-{\,}{1\over 2}{\,}R{\,}g_{\mu\nu}{\quad};
\eqno (2.13)$$
Following a lengthy calculation, one finds [3] that the Einstein field
equations, up to and including second order in perturbations, can be
written as
$$G^{(0)}_{\mu\nu}{\;}{\,}
={\;}{\,}8\pi{\,}T^{(0)}_{\mu\nu}{\,}+{\,}8\pi{\,}T^{(2)}_{\mu\nu}{\,} 
+{\,}8\pi{\,}T_{\mu\nu}^{'}{\,} 
-{\,}G^{(1)}_{\mu\nu}{\quad},\eqno(2.14)$$
\noindent
Here, $T^{(0)}_{\mu\nu}$ is the background energy-momentum tensor
(2.5), and $G^{(1)}_{\mu\nu}$ denotes ${\,}-({1\over 2})\times$ the 
left-hand side of Eq.(2.6).  The quantity $T^{(2)}_{\mu\nu}$ denotes
$$\eqalign{T^{(2)}_{\mu\nu}{\;}{\,} 
={\;}{\,}\nabla_{\mu}\phi^{(1)}{\,}&\nabla_{\nu}\phi^{(1)}
-{1\over 2}{\,}\gamma_{\mu\nu}{\,}\gamma^{\rho\sigma}{\,}
{\,}\nabla_{\rho}\phi^{(1)}{\,}\nabla_{\sigma}\phi^{(1)}
+\Bigl(\gamma_{\mu\nu}{\,}h^{(1)\sigma\rho}
-h^{(1)}_{\mu\nu}{\,}\gamma^{\sigma\rho}\Bigr)
\nabla_{\sigma}\Phi{\;}\nabla_{\rho}\phi^{(1)} \cr
&+{\,}{1\over 2}{\,}\Bigl(h^{(1)}_{\mu\nu}{\,}h^{(1)\sigma\rho}
-\gamma_{\mu\nu}{\,}h^{(1)\sigma\alpha}{\,}h^{(1)}_{\alpha}{}^{\rho}\Bigr)
\nabla_{\sigma}\Phi{\;}\nabla_{\rho}\Phi\cr}\eqno(2.15)$$ 
\noindent
and $T'_{\mu\nu}$ is defined by
$$\eqalign{8\pi{\,}T'_{\mu\nu}{\;}{\,}
&={\;}{\,}{1\over 4}{\,}
\Bigl(\bar{h}^{(1)\sigma\rho}{}_{;\mu}{\;}h^{(1)}_{\sigma\rho}{}_{;\nu}{\,}
-{\,}2{\,}\bar{h}^{(1)}_{\alpha\sigma}{}^{;\alpha}{\;}
\bar{h}^{(1)\sigma}{}_{(\mu;\nu)}\Bigr)
-{1 \over 2}{\,}\bar{h}^{(1)\sigma}{}_{(\mu}R^{(0)}_{\nu)
\rho\sigma\alpha}\bar{h}^{(1)\alpha\rho}\cr
&+{1 \over 2}{\,}\bar{h}^{(1)}_{\sigma(\mu}{\,}
R^{(0)}_{\nu)\alpha}{\,}\bar{h}^{(1)\alpha\sigma}
-{1 \over 2}{\,}h^{(1)\sigma}{}_{(\mu}\bar{h}^{(1)}_{\nu)\sigma}{\,}R^{(0)}
-{\,}8\pi{\,}T_{\sigma(\mu}^{(1)}\bar{h}^{(1)}_{\nu)}{}^{\sigma}\cr
&-4\pi{\,}\gamma_{\mu\nu}{\,}
\Bigl(2{\,}\bar{h}^{(1)\sigma\rho}{\,}
\nabla_{\sigma}\phi^{(1)}{\,}\nabla_{\rho}\Phi{\,}
+{\,}\phi^{(1)}{\,}\nabla_{\sigma}\nabla^{\sigma}\phi^{(1)}
-{\,}\bar{h}^{(1)\sigma\rho}{\,}h^{(1)}_{\sigma}{}^{\beta}{\,}
\nabla_{\rho}\Phi{\,}\nabla_{\beta}\Phi\Bigr)\cr
&+{\,}C^{\sigma}_{\mu\nu ;\sigma}{\quad},\cr}\eqno (2.16)$$
\noindent
where the explicit form of $C^{\sigma}_{\mu\nu}$ will not be needed.\par 
\smallskip
\indent
In the high-frequency limit, after Brill-Hartle or Isaacson averaging
[9] over many wavelengths (both in space and in time), as summarised
in [3], $<T'_{\mu\nu}>$ will give the leading spin-2 (graviton)
contribution to $G^{(0)}_{\mu\nu}{\,}$, and $T^{(2)}_{\mu\nu}$ will
give the contribution quadratic in the scalar fluctuations 
${\phi}^{(1)}{\,}$.  On averaging over many wave periods and over 
angles, following Sec.3 of [3], one finds for late Lorentzian times 
in the high-frequency (Isaacson) approximation:
$$<{T'}_{\mu\nu}>{\;}{\,}
={\;}{\,}{1 \over 32\pi}{\,} 
<{\bar h}^{(1)\sigma\rho}_{~~~~~;\mu}{\;}h^{(1)}_{\sigma\rho;\nu}{\,} 
-{\,}2{\;}{\bar h}^{(1);\alpha}_{\alpha\sigma} \;
{\bar h}^{(1)\sigma}_{~~~~(\mu;\nu)}>{\quad}.\eqno(2.17)$$ 
In the transverse-traceless gauge (2.9,11), appropriate for this
region of the space-time, Eq.(2.3) simplifies to give
$$T^{GW}_{\mu\nu}{\;}{\,} 
\equiv{\;}{\,}<{T'}_{\mu\nu}>_{TT}{\;}{\,}
={\;}{\,}{1\over 32\pi}{\;}<{h^{(1)\sigma\rho}}_{;\mu}{\;}
h^{(1)}_{\sigma\rho;\nu}>_{TT}{\quad},\eqno(2.18)$$
\indent
The Isaacson averaged energy-momentum contribution of the scalar-field
fluctuations, taken for simplicity in the late-time region where the
background scalar field $\Phi$ is nearly zero, is 
$$<T^{(2)}_{\mu\nu}>{\;}{\,}
={\;}{\,}<\nabla_{\mu}\phi^{(1)}{\;}
\nabla_{\nu}\phi^{(1)}>{\quad}.\eqno(2.19)$$ 
\smallskip
\indent
Spin-1 Maxwell field perturbations can also be treated in a similar
way [16].  For a perturbative Maxwell vector potential $A^{(1)}_{\mu}$ 
in the Lorentz gauge
$$\nabla^{\mu}A^{(1)}_{\mu}{\;}{\,}={\;}{\,}0{\quad},\eqno(2.20)$$
\noindent
the Maxwell field equations read [9]
$$\nabla^{\mu}\nabla_{\mu}A^{(1)}_{\nu}{\,}
-{\,}R^{(0)}_{\mu\nu}{\;}A^{(1)\mu}{\;}{\,}={\;}{\,}0{\quad}.\eqno(2.21)$$  
\noindent
The averaged Maxwell energy-momentum tensor
$$<T^{\alpha\beta}>_{\rm Maxwell}{\;}{\,} 
={\;}{\,}{1\over 4\pi}{\;}<\nabla^{\alpha}A^{(1)\sigma}{\,} 
\nabla^{\beta}A^{(1)}_{\sigma}{\;}
+{\;}{R^{(0)\alpha}}_{\nu}{\;}A^{(1)\nu}{\,}A^{(1)\beta}>_{\rm Lor}
\eqno(2.22)$$
\noindent
in Lorentz gauge can be simplified using the field equation (2.21) and
integration by parts in the Isaacson limit, to give
$$<T^{\alpha\beta}>_{\rm Maxwell}{\;}{\,} 
={\;}{\,}{1\over 4\pi}{\;}<\nabla_{\mu}A^{(1)\alpha}{\;}
\nabla_{\nu}A^{(1)}_{\alpha}>_{\rm Lor}{\quad}.\eqno(2.23)$$ 
\noindent
For spin-1 Yang-Mills fields, which typically appear when working with
locally-supersymmetric theories of supergravity coupled to supermatter 
[10], a similar but more complicated treatment can be given.\par
\smallskip
\indent
Thus Eq.(2.14), averaged over high-frequency fluctuations, and
including a spin-1 Maxwell-field contribution, becomes 
$$\eqalign{G^{(0)}_{\mu\nu}(\gamma){\;}{\,}
={\;}{\,}8\pi <\nabla_{\mu}\phi^{(1)}{\,}&\nabla_{\nu}\phi^{(1)}>{\;}
+{\;}2 <\nabla_{\mu}A^{(1)\alpha}{\;}
\nabla_{\nu}A^{(1)}_{\alpha}>_{\rm Lor}\cr
&+{\;}{1\over 4} <{h^{(1)\sigma\rho}}_{;\mu}{\;}
{h^{(1)}}_{\sigma\rho;\nu}>_{TT}{\quad}.\cr}\eqno(2.24)$$
\noindent
Further perturbative corrections to Eq.(2.24) are of a relative size
$O(\epsilon)$ smaller, and it must be understood that one solves Eqs.
(2.12,21,24) simultaneously.  To ease the notation, we henceforth drop 
the labels $\rm Lor$ and $TT$.\par
\smallskip
\indent
For high-frequency (real) massless perturbations 
$\phi^{(1)},A^{(1)}_{\mu},h^{(1)}_{\mu\nu}{\,}$, we make an Ansatz 
which is natural for late times:
$$\eqalignno{\phi^{(1)}(x){\;}{\,}& 
={\;}{\,}\sum^{\infty}_{\ell =0}{\;}\sum^{\ell}_{m=-\ell}{\;}{\,}
\int^{\infty}_{0}{\;}d\omega{\;}{\,}\Bigl[A_{\omega\ell m}(t,r,\Omega){\;}
{e^{i\theta{_\omega}(t,r)/\epsilon}}{\;}+{\,}{\rm c.c.}\Bigr]
{\quad},&(2.25)\cr 
A^{(1)}_{\mu}(x){\;}{\,}&
={\;}{\,}\sum^{\infty}_{\ell =1}{\;}\sum^{\ell}_{m=-\ell}{\;}\sum_{P}{\;} 
\int^{\infty}_{0}{\;}d\omega{\;}{\,}
\Bigl[(A_{\mu})_{\omega\ell mP}(t,r,\Omega){\;}
e^{i\theta_{\omega}(t,r)/\epsilon}{\,}+{\;}{\rm c.c.}\Bigr]{\quad},&(2.26)\cr
h^{(1)}_{\mu\nu}(x){\;}{\,}&
={\;}{\,}\sum^{\infty}_{\ell =2}{\;}\sum^{\ell}_{m=-\ell}{\;}\sum_{P}{\;} 
\int^{\infty}_{0}{\;}d\omega{\;}{\,}
\Bigl[(A_{\mu\nu})_{\omega\ell mP}(t,r,\Omega){\;}
e^{i\theta_{\omega}(t,r)/\epsilon}{\;}
+{\;}{\rm c.c.}\Bigr]{\quad},&(2.27)\cr}$$
\noindent
where $P=+{\,},{\,}\times{\,}$ denotes the two orthogonal polarisation 
states for a radially-travelling gravitational wave, and also the 
standard independent polarisation states in the $\theta$-and 
$\phi$-directions for a radially-travelling electromagnetic wave [9].
The quantity $\theta_{\omega}(t,r)/\epsilon$ is a rapidly-varying real 
phase, in common to all the spins 0, 1 and 2, which precisely allows
for the predominantly radial wave-propagation at late times.  Defining 
$\psi =\theta_{\omega}(t,r)/\epsilon{\,}$, we assume that the first 
derivative of $\psi$ is large in comparison with first derivatives of 
the 'amplitude' $A_{\omega\ell m}$ or of the corresponding tetrad 
components of $(A_{\mu})_{\omega\ell mP}$ or
$(A_{\mu\nu})_{\omega\ell mP}{\,}$  -- see below.  Schematically, 
${\;}{\mid}\partial\psi/\psi{\mid}{\;}\gg{\;}{\mid}\partial A/A{\mid}
{\quad}$.\par
\smallskip
\indent
One might expect the high-frequency expansions (2.25-27) for the
perturbed scalar, Maxwell and graviton fields to give a good
approximation to the radially-outgoing radiation at late times during
the 'long' period when the black hole is radiating in a quasi-static
way, with its mass 'slowly' decreasing from the initial value $M_I{\,}$.
The corresponding approximate behaviour of the overall 
spherically-symmetric 'background' gravitational field at late times 
is expected to be given by the Vaidya metric [7], as treated in Sec.3 
below.  In particular, the connection between the late-time 
high-frequency expansions (2.25-27) and the Vaidya metric will be 
described explicitly in Sec.3.  Of course, in our classical 
boundary-value formulation, with the time-at-infinity $T$ taken to be 
of the form ${\,}T={\mid}T{\mid}e^{-i\theta}{\,}$, for
$0<\theta\leq\pi/2{\,}$, the amplitudes 
$A_{\omega\ell m},{\;}(A_{\mu})_{\omega\ell mP}$ and 
$(A_{\mu\nu})_{\omega\ell mP}$ in the late-time high-frequency 
expansions (2.25-27) are related to the scalar, spin-1 Maxwell and 
spin-2 graviton data on the final surface $\Sigma_F{\,}$, with time 
$T$ at infinity.  Conversely, on following the radiation to the past, 
we reach the strongly-interacting collapse region of the space-time, 
where both the background spherically-symmetric metric
$\gamma_{\mu\nu}$ and the scalar field $\Phi$ may vary rapidly with the 
coordinates $t$ and $r{\,}$.  It is only because the background 
$(\gamma_{\mu\nu}{\,},\Phi)$ is the {\it complex} solution of the 
boundary-value problem for the spherically-symmetric Einstein/scalar 
system, with a {\it complex} time-separation-at-infinity 
${\,}T={\mid}T{\mid}\exp(-i\theta)$, for ${\,}0<\theta<\pi/2{\,}$, 
that large deviations from flatness in the boundary data are expected 
to be smoothed out in the usual elliptic fashion.  This is the 
distinguishing feature of this complex approach; in contrast, 
Lorentzian-signature evolution of the Einstein field equations, 
including matter, generically leads to space-time singularities.  If 
one knew the form of the background solution, then (computationally, 
at least) one could solve the coupled evolution equations for
harmonics of (say) the perturbed scalar and gravitational fields.  One 
would then, by matching of asymptotic expansions [11], have to join the
wave-like solutions emerging from the strong-field 'collapse region'
above onto the high-frequency expansions (2.25-27) for the radiative 
parts of the fields at late times.\par
\smallskip
\indent
Next, we consider the leading, geometrical-optics limit of the
perturbative field equations for spin-0, 1, and 2.  For spin-0, one has 
$$\gamma^{\mu\nu}{\;}\phi^{(1)}_{;\mu\nu}{\;}{\,}
={\;}{\,}0{\quad},\eqno(2.28)$$
\noindent
which is the $O(\epsilon^{1})$ part of the scalar field equation
Eq.(2.4), in the late-time limit that $\Phi =0{\,}$.  The spin-1 field
equations are given in Eq.(2.21), and the linearised spin-2 field equations 
in Eq.(2.10).  Define
$$(k_{\mu})_\omega{\;}{\,}={\;}{\,}\nabla_{\mu}\theta_{\omega}
{\quad}.\eqno(2.29)$$
\noindent
Working again in the late-time region and taking $\Phi =0$ there, a
straightforward calculation, applying the perturbative field equations
(2.10,21,28) to the high-frequency expansions (2.25-27), together with 
the $TT$ and Lorentz gauge conditions, leads to the following properties:
$$\eqalignno{(k^{\mu})_{\omega}{\;}(k_{\mu})_{\omega}{\;}{\,}& 
={\;}{\,}0{\quad},&(2.30)\cr 
(A_{\mu})_{\omega\ell mP}{\;}(k^{\mu})_{\omega}{\;}{\,}&
={\;}{\,}0{\quad},&(2.31)\cr
(A_{\mu\nu})_{\omega\ell mP}{\;}(k^{\mu})_{\omega}{\;}{\,}&
={\;}{\,}0{\quad},&(2.32)\cr
\gamma^{\mu\nu}{\;}(A_{\mu\nu})_{\omega\ell mP}{\;}{\,}&
={\;}{\,}0{\quad},&(2.33)\cr}$$
\noindent
at lowest order in $\epsilon{\,}$.  Suppose that the background metric
$\gamma_{\mu\nu}$ in this region, written with respect to coordinates
$(t,r,\theta,\phi)$ in the form
$$ds^{2}{\;}{\,}
={\;}{\,}-{\;}e^{b(t,r)}{\;}dt^{2}{\,}+{\;}e^{a(t,r)}{\;}dr^{2}{\;} 
+{\;}r^{2}{\,}\Bigl(d\theta^{2}{\,}+{\,}\sin^{2}\theta{\,}d\phi^{2}\Bigr)
{\quad},\eqno(2.34)$$
\noindent
has the Vaidya form [7], but in the coordinate system described in
Eq.(3.18) below:
$$e^{-a(t,r)}{\;}{\,}={\;}{\,}1{\,}-{\,}{{2m(t,r)}\over{r}}{\quad},{\qquad}
e^{b(t,r)}{\;}{\,}={\;}{\,}\biggl({{\dot m}\over{f(m)}}\biggr)^{2}{\;}e^{-a}
{\quad}.\eqno(2.35)$$
\noindent
Here, $m(t,r)$ is a slowly-varying 'mass function', with 
$\dot m=(\partial  m/\partial t)$, and where $f(m)$ depends on the 
details of the radiation.  Then, for each choice of the integration 
variable $\omega$ in Eqs.(2.25-27), Eq.(2.29) has an outgoing-wave solution
$$\theta_{\omega}(t,r){\;}{\,} 
={\;}{\,}\omega{\,}(t{\,}-{\,}r^{*}){\quad},\eqno(2.36)$$
\noindent
where we define
$$r^{*}{\;}{\,} 
={\;}{\,}\int^{r}d{\hat r}{\;}e^{a(t,{\hat r})}{\quad},\eqno(2.37)$$
\noindent
by analogy with the 'tortoise coordinate' 
${\,}r^{*}{\,}={\,}r{\,}+{\,}2M\ln\Bigl((r/2M)-1\Bigr)$ in the 
Schwarzschild solution [9,12].  Because of the slowly-varying nature
of the background, one has 
${\,}{\mid}\partial_{t}r^{*}{\mid}\ll 1{\,}$.  Note that a general 
solution of Eq.(2.30): 
$(k^{\mu})_{\omega}{\,}(k_{\mu})_{\omega}=0{\,}$, would involve a 
general function of ${\,}(t{\,}\pm{\,}r^{*})$.  The outgoing-wave 
solution (2.36) is picked out because we require the expansions 
(2.25-27) to reduce to outgoing Fourier expansions at large radius.\par
\smallskip
\indent
In a standard fashion, the application of the linearised field
equations and gauge conditions to the high-frequency expansions 
(2.25-27) can be carried on to the next order, one power of $\epsilon$
beyond geometrical optics. For the spin-0 perturbations, one finds
$$A_{\omega\ell m}{\;}\nabla^{\mu}(k_{\mu})_{\omega}{\;} 
+{\;}2{\,}(k^{\mu})_{\omega}{\;}\nabla_{\mu}A_{\omega\ell m}{\;}{\,} 
={\;}{\,}0{\quad},\eqno(2.38)$$
\noindent
whence
$$\nabla^{\mu}\Bigl[{\mid}A_{\omega\ell m}{\mid}^{2}{\;}
(k_{\mu})_{\omega}\Bigr]{\;}{\,} 
={\;}{\,}0{\quad}.\eqno(2.39)$$
\indent
For the spin-1 field, one finds
$$(A_{\nu})_{\omega\ell mP}{\;}\nabla^{\sigma}(k_{\sigma})_{\omega}{\;}
+{\;}2{\,}(k^{\sigma})_{\omega}{\;}
\nabla_{\sigma}(A_{\nu})_{\omega\ell mP}{\;}{\,}
={\;}{\,}0{\quad}.\eqno(2.40)$$
\noindent
Now introduce a polarisation vector $(e_{\mu})_{\omega\ell mP}$ 
such that
$$(A_{\mu})_{\omega\ell m}{\;}{\,}={\;}{\,}A_{1\omega\ell mP}{\;}
(e_{\mu})_{\omega\ell mP}{\quad},\eqno(2.41)$$
$$(e_{\mu})_{\omega\ell mP}{\;}(e^{\mu})^{*}_{\omega\ell mP'}{\;}{\,} 
={\;}{\,}\delta_{PP'}{\quad},\eqno(2.42)$$ 
$$A_{1\omega\ell mP}{\;}{\,}
={\;}{\,}\Bigl[(A_{\mu})_{\omega\ell mP}{\;}
(A^{\mu})^{*}_{\omega\ell mP}\Bigr]^{1\over 2}{\quad},\eqno(2.43)$$
\noindent
where a star denotes complex conjugation.  The Lorentz condition implies
$$(e^{\mu})_{\omega\ell mP}{\;}(k_{\mu})_{\omega}{\;}{\,} 
={\;}{\,}0{\quad}.\eqno(2.44)$$
\noindent
The Maxwell field equations then imply
$$\nabla^{\mu}\Bigl[{\mid}A_{1\omega\ell mP}{\mid}^{2}
{\;}(k_{\mu})_{\omega}\Bigr]{\;}={\;}0{\;}.\eqno(2.45)$$
\smallskip
\indent
Correspondingly, for the spin-2 field (gravitons), one finds 
$$(A_{\mu\nu})_{\omega\ell mP}{\;}{(k_{\sigma})_{\omega}}^{;\sigma}{\;} 
+{\;}2{\,}(k^{\sigma})_{\omega}{\;}
(A_{\mu\nu})_{\omega\ell mP;\sigma}{\;}{\,}
={\;}{\,}0{\quad}.\eqno(2.46)$$
\noindent
One then introduces a symmetric polarisation tensor
$(e_{\mu\nu})_{\omega\ell mP}$ such that
$$(A_{\mu\nu})_{\omega\ell mP}{\;}{\,} 
={\;}{\,}A_{2\omega\ell mP}{\;}(e_{\mu\nu})_{\omega \ell mP}{\quad},
\eqno(2.47)$$ 
$$(e_{\mu\nu})_{\omega\ell mP}{\;}(e^{\mu\nu})^{*}_{\omega\ell mP'}{\;}{\,} 
={\;}{\,}\delta_{PP'}{\quad},\eqno(2.48)$$
$$A_{2\omega\ell mP}{\;}{\,}
={\;}{\,}\Bigl[(A_{\mu\nu})_{\omega\ell mP}{\;}
(A^{\mu\nu})^{*}_{\omega\ell mP}\Bigr]^{1\over2}{\quad},\eqno(2.49)$$
\noindent
where the last equality is valid up to an unimportant phase. The $TT$
condition implies
$$\eqalignno{(e_{\mu\nu})_{\omega\ell mP}{\;}(k^{\mu})_{\omega}{\;}{\,}&
={\;}{\,}0{\quad},&(2.50)\cr
\gamma^{\mu\nu}{\;}(e_{\mu\nu})_{\omega\ell mP}{\;}{\,}&
={\;}{\,}0{\quad}.&(2.51)\cr}$$
\noindent
Then the linearised spin-2 field equations imply
$$\nabla^{\mu}\Bigl[{\mid}A_{2\omega\ell mP}{\mid}^{2}{\;}
(k_{\mu})_{\omega}\Bigr]{\;}{\,}={\;}{\,}0{\quad}.\eqno(2.52)$$
\indent
For $s = 1$ and $2{\,}$, write
$$A_{s\ell m\omega P}{\;}{\,}
={\;}{\,}{\mid}A_{s\omega\ell m\omega P}{\mid}{\;}
e^{i\sigma_{s\omega\ell mP}}{\quad},\eqno(2.53)$$
\noindent
where $\sigma_{s\omega\ell mP}$ is a real phase.  From the 'evolution
equations' (2.45,52), one finds that
$$(k^{\mu})_{\omega}{\;}\nabla_{\mu}\sigma_{s\omega\ell mP}{\;}{\,} 
={\;}{\,}0{\quad},\eqno(2.54)$$
\noindent
provided that
$$(k^{\mu})_{\omega}{\;}
\nabla_{\mu}\Bigl(\ln{\mid}A_{s\omega\ell mP}{\mid}\Bigr){\;}{\,} 
={\;}{\,}-{\;}{1\over 2}{\;}\nabla^{\mu}(k_{\mu})_{\omega}
{\quad}.\eqno(2.55)$$
\noindent
A similar equation holds for the evolution of the spin-0 coefficients
$A_{\omega\ell m}{\,}$.  Now define a preferred affine parameter 
$\lambda$ along the null rays such that
$$(k^{\mu})_{\omega}{\;}{\,} 
={\;}{\,}{{dx^{\mu}}\over{d\lambda}}{\quad},\eqno(2.56)$$ 
\noindent
and the $x^{\mu}(\lambda)$ are affinely parametrised null geodesics.  
This can alternatively be written in the form
$$(k_{\mu})_{\omega}{\;}\nabla^{\mu}(k_{\nu})_{\omega}{\;}{\,} 
={\;}{\,}0{\quad};\eqno(2.57)$$
\noindent
that is, that the high-frequency waves move along null geodesics.
From Eq.(2.55), one sees that the amplitude 
${\mid}A_{s\omega\ell mP}{\mid}$ decreases if 
${\,}\nabla^{\mu}(k_{\mu})_{\omega}>0{\,}$, that is, if the null rays 
diverge.  In the arguments leading to Eqs.(2.45,53), one finds also 
that the corresponding polarisation tensors are parallely transported 
along the null geodesic $x^{\mu}(\lambda){\,}$.  That is,
$$\eqalignno{(k^{\sigma})_{\omega}{\;}\nabla_{\sigma}
(e_{\mu})_{\omega\ell mP}{\;}{\,}&
={\;}{\,}0{\quad},&(2.58)\cr
(k^{\sigma})_{\omega}{\;}
\nabla_{\sigma}(e_{\mu\nu})_{\omega\ell mP}{\;}{\,}&
={\;}{\,}0{\quad}.&(2.59)\cr}$$
\noindent
Of course, such a description in terms of a family of null geodesics
will only be valid in a comparatively late-time, large-distance region
of the space-time.  Where space-time becomes highly curved, caustics
would be expected to develop and the geometrical-optics approach would
break down.\par
\smallskip
\indent
Turning again to the Einstein field equations, we calculate the
quantities on the right-hand side of Eq.(2.24), being the contributions
to $G^{(0)}_{\mu\nu}(\gamma)$ which are quadratic in the spin-0,
spin-1, and spin-2 fluctuations.  As earlier, $<  >$ denotes an
Isaacson average over times and angles.  For an incoherent source of
waves, comprising a large number of roughly stationary sources
(essential for near-spherical symmetry), only terms in Eqs.(2.26,27)
with $\ell =\ell'{\,},{\,}m=m'{\,}$ contribute to the average.  In the
context of Eqs.(2.26,27), $< >$ is also an average over the random
phase ${\,}\theta_\omega{\,}$, since a time average.  Therefore, 
at leading order $O(\epsilon^{-2})$ with respect to the high-frequency
approximations (2.25-27), one has
$$<\nabla_{\mu}\phi^{(1)}{\;}\nabla_{\nu}\phi^{(1)}>{\;}{\,}
={\;}{\,}{{2}\over{\epsilon^{2}}}{\;}\sum_{\ell m}{\;}\int^{\infty}_{0}{\;}
d\omega{\;}{\,}(k_{\mu})_{\omega}{\;}(k_{\nu})_{\omega}{\;}
{\mid}A_{\omega\ell m}(t,r){\mid}^{2}{\quad},\eqno(2.60)$$
\noindent
and
$$<\nabla_{\mu}A^{\sigma}{\;}\nabla_{\nu}A_{\sigma}>{\;}{\,}
={\;}{\,}{{2}\over{\epsilon^{2}}}{\;}\sum_{\ell mP}{\;}\int^{\infty}_{0}{\;}
d\omega{\;}{\,}(k_{\mu})_{\omega}{\;}(k_{\nu})_{\omega}{\;}
{\mid}A_{1\omega\ell mP}(t,r){\mid}^{2}{\quad}.\eqno(2.61)$$
\noindent
Further,
$$<\nabla_{\mu}h^{(1)}_{\sigma\rho}{\;}
\nabla_{\nu}h^{(1)\sigma\rho}>{\;}{\,}
={\;}{\,}{{2}\over{\epsilon^{2}}}{\;}\sum_{\ell mP}{\;}\int^{\infty}_{0}{\;}
d\omega{\;}{\,}(k_{\mu})_{\omega}{\;}(k_{\nu})_{\omega}{\;}
{\mid}A_{2\omega\ell mP}(t,r){\mid}^{2}{\quad}.\eqno(2.62)$$
\noindent
Here, we define the quantity ${\mid}A_{s\omega\ell mP}(t,r){\mid}^2$ 
to be
$${\mid}A_{s\omega\ell mP}(t,r){\mid}^{2}{\;}{\,} 
={\;}{\,}{{1}\over{4\pi}}{\;}\int{\,}d\Omega{\;}
<{\mid}A_{s\omega\ell mP}(x){\mid}^{2}>_{\theta}{\quad},\eqno(2.63)$$
\noindent
where $<  >_\theta$ denotes a time or phase average.  Define further
$$<T_{\mu\nu}>{\;}{\,}
={\;}{\,}<T^{(2)}_{\mu\nu}>{\,}
+{\,}<T_{\mu\nu}>_{{\rm Maxwell}}{\,}
+{\,}T^{GW}_{\mu\nu}{\quad}.\eqno(2.64)$$
\noindent
Combining the high-frequency approximation with Isaacson averaging
leads to the tensor (at leading order)
$$<T_{\mu\nu}>{\;}{\,}
={\;}{\,}{{2}\over{\epsilon^{2}}}{\;}\sum_{s\ell mP}{\;}c_{s}{\;}
\int^{\infty}_{0}{\;}d\omega{\;}{\,}(k_{\mu})_{\omega}{\;}
(k_{\nu})_{\omega}{\;}{\,}{\mid}A_{s\omega\ell mP}(t,r){\mid}^{2}{\quad},
\eqno(2.65)$$ 
\noindent
where
$$c_{0}{\;}{\,}={\;}{\,}1{\quad},
{\qquad}c_{1}{\;}{\,}={\;}{\,}{{1}\over{4\pi}}{\quad},
{\qquad}c_{2}{\;}{\,}={\;}{\,}{{1}\over{32\pi}}{\quad}.\eqno(2.66)$$
\noindent
As one would expect for a null fluid, one has
$$<{T^{\sigma}}_{\sigma}>{\;}{\,}={\;}{\,}0\eqno(2.67)$$
\noindent
to leading order.  The quantity $\epsilon$ was regarded as a free
parameter above, which helps in keeping track of the magnitudes of
different quantities in the high-frequency approximation.  But, given
that the terms denoted by 
$\exp\bigl(i\theta_{\omega}(t,r)/\epsilon\bigr)$ in Eqs.(2.25-27) are 
indeed of high frequency, one may then set $\epsilon =1$ in future 
calculations, without loss of generality.\par
\smallskip
\indent
One can readily show that $<T_{\mu\nu}>$ transforms as a tensor under
$(t,r)$-dependent 'background' coordinate transformations.  Further,
the equations of continuity (2.39,45,52) imply the conservation
equation ${\,}\nabla^{\nu}<T_{\mu\nu}>=0{\,}$.  It is then natural 
to regard ${\,}c_{s}{\,}{\mid}A_{s\omega\ell mP}{\mid}^{2}$, 
the total intensity in the high-frequency perturbations, as a measure 
of the total energy density.  But while the total energy is
independent of the choice of space-like hypersurface, the notion of 
energy density only has significance with respect to a particular 
space-like hypersurface.  Denoting the unit future-directed time-like 
normal vector to the hypersurface by $n^{(0)\mu}{\,}$, the energy 
density measured locally by an observer with 4-velocity $n^{(0)\mu}$ is
$$\eqalign{\rho{\;}{\,}&
={\;}{\,}n^{(0)\mu}{\;}n^{(0)\nu}{\,}<T_{\mu\nu}>\cr
&={\;}{\,}2\sum_{s\ell mP}{\;}\int^{\infty}_{0}{\,}d\omega{\;}{\,}
(n.k_{\omega})^{2}{\;}c_{s}{\;}
{\mid}A_{s\omega\ell mP}(t,r){\mid}^{2}{\quad}.\cr}\eqno(2.68)$$
\noindent
As expected in perturbation theory about the spherically-symmetric
background $(\gamma_{\mu\nu}{\,},\Phi)$, the mass-energy of the massless
waves is quadratic in their amplitude, for small 
${\mid}A_{s\omega\ell mP}{\mid}$.  A further consequence of 
Eqs.(2.39,45,52) is that the quantity
$$N_{\omega}{\;}{\,} 
={\;}{\,}\sum_{s\ell mP}{\;}\int_{\Sigma}{\;}d^{3}x{\;}{\,}
{\sqrt{^{(3)}\gamma}}{\;}{\,}(n_{.}k_{\omega}){\;}
{\mid}A_{s\omega\ell mP}{\mid}^{2}\eqno(2.69)$$ 
\noindent
is the conserved total number density (independent of space-like
hypersurface) of massless waves (massless-scalar particles, photons
and gravitons) passing through the space-like hypersurface 
$\Sigma{\,}$.\par
\medskip
\noindent
{\bf 3. Solution of background field equations}
\medskip
\indent
The Einstein field equations for a spherically-symmetric geometry of
Lorentzian signature, of the form (2.34), may be derived from the
Riemannian field equations, as given in Eq.(3.5-11) of [3], on
replacing $e^b$ by $(-e^{b}){\,}$:
$$\eqalignno{a'{\;}{\,}&
={\;}{\,}8\pi{\,}r{\,}T_{rr}{\;}+{\;}{(1-e^{a})\over r}{\quad},&(3.1)\cr 
b'{\;}{\,}&
={\;}{\,}8\pi{\,}r{\,}e^{a-b}{\;}T_{tt}{\;}
-{\;}{(1-e^{a})\over r}{\quad},&(3.2)\cr
{\dot a}{\;}{\,}&
={\;}{\,}8\pi{\,}r{\,}T_{tr}{\quad},&(3.3)\cr}$$
$$1-{\,}e^{-a}{\,}+{\,}{1\over 2}{\;}r{\,}e^{-a}{\,}(a'-b'){\;} 
-{\,}{1\over 2}{\;}r^{2}{\;}R^{(0)}{\;}{\,}
={\;}{\,}8\pi{\,}T_{\theta\theta}{\;}{\,} 
={\;}{\,}{{8\pi{\,}T_{\phi\phi}}\over{\sin^{2}\theta}}{\quad},\eqno(3.4)$$
\noindent
where
$$R^{(0)}{\;}{\,} 
={\;}{\,}-{{2}\over{r^{2}}}{\,}\bigl(1-e^{-a}\bigr){\;}
+{\;}e^{-{1\over 2}(a+b)}{\;}
\biggl[\partial_{t}\Bigl({\dot a}{\,}e^{{1\over 2}(a-b)}\Bigr){\,} 
-{\,}\partial_{r}\bigl(b'{\,}e^{{1\over 2}(b-a)}\bigr)\biggr]{\;}{\,}
={\;}{\,}-{\,}8\pi{\,}{T^{\mu}}_{\mu}{\quad}.\eqno(3.5)$$
\smallskip
\noindent
As usual, for a massless scalar field ${\,}\Phi(t,r)$, one has
${\,}T_{\mu\nu}{\,}={\,}\Phi_{,\mu}\Phi_{,\nu}-{1 \over 2}{\,}
\gamma_{\mu\nu}{\,}(\Phi_{,\alpha}\Phi_{,\beta}{\,}
\gamma^{\alpha\beta})$.\par
\smallskip
\indent
Equations (3.4,5) imply that
$$T_{rr}{\;}{\,}={\;}{\,}e^{(a-b)}{\;}T_{tt}{\quad},\eqno(3.6)$$
\noindent
whence, by Eqs.(3.1,2),
$${1\over 2}{\,}(a'-b'){\;}{\,}={\;}{\,}{{(1-e^{a})}\over r}{\quad}.
\eqno(3.7)$$
\noindent
We now derive the Vaidya metric, corresponding to a
spherically-symmetric null-fluid source, in the form (2.34,35) quoted
above.  Taking the metric form (2.34), we {\it define} the function
$m(t,r)$ by
$$e^{-a(t,r)}{\;}{\,}={\;}{\,}1{\,}-{\,}{{2m(t,r)}\over r}{\quad}.
\eqno(3.8)$$
\noindent
Using Eq.(3.7), we deduce an expression for ${\,}e^b(t,r){\,}$:
$$e^{b(t,r)}{\;}{\,} 
={\;}{\,}\biggl(1-{{2m(t,r)}\over r}\biggr){\;}
\exp\biggl[4\int^{r}_{{\hat r}}{\,}d{\bar r}{\;}{\,}
{{m'(t,{\bar r})}\over{({\bar r}-2m(t,{\bar r}))}}\biggr]{\quad},\eqno(3.9)$$
\noindent
for some ${\hat r}{\,}$.  By elementary flatness at the origin, 
one must have $a\rightarrow 0$ as $r\rightarrow 0$ on each space-like
hypersurface.  Asymptotic flatness requires setting 
${\hat r}=R_{\infty}$ at the outer boundary, and then taking the limit
$R_{\infty}\rightarrow\infty{\,}$.  Eq.(3.1) is the Hamiltonian
constraint equation [9,13].  Use of Eq.(3.8) shows that Eq.(3.1) can
be written as a first-order differential equation for the mass $m(r)$ 
inside a radius $r$ at time $t=t_{0}{\,}$, say:
$${{\partial m}\over{\partial r}}{\;}{\,}
={\;}{\,}4\pi{\,}r^{2}{\,}\rho{\quad},\eqno(3.10)$$
\noindent
where
$$\rho{\;}{\,}={\;}{\,}e^{-b}{\;}T_{tt}\eqno(3.11)$$
\noindent
is the energy density.  Eq.(3.3) is the momentum constraint equation
[9,13].\par
\smallskip
\indent
We can now determine the background metric at late times, when the
energy-momentum tensor is that of the black-hole radiation, following
Vaidya [7].  As in Sec.2, we study the gravitational field produced
by perturbations whose averaged energy-momentum tensor is
$<T_{\mu\nu}>$.  (Later, we shall move to a coordinate system more
suited to retarded radiation.)  Since the direction 
${\,}(k^\mu)_{\omega}$ of propagation of the radiation in Sec.2 is
null, we choose
$$(k_{r})_{\omega}{\;}e^{{1\over 2}(b-a)}{\;}
+{\;}(k_{t})_{\omega}{\;}{\,}={\;}{\,}0{\quad},\eqno(3.12)$$
\noindent
which corresponds to an outgoing-wave boundary condition at large $r$.  
Eq.(3.3) also implies
$$<T^{t}_{~r}>{\,}e^{{1\over 2}(b-a)}{\;}
+{\;}<{T^{t}}_{t}>{\;}{\,}={\;}{\,}0{\quad}.\eqno(3.13)$$
\noindent
The field equations in terms of the metric functions $a$ and $b$ are
as in Eqs.(3.1-5), but with $T_{\mu\nu}$ replaced by $<T_{\mu\nu}>$.  
Using the momentum constraint (3.3), the Hamiltonian constraint (3.1), 
with Eq.(3.13) and ${\,}\rho =-<{T^{t}}_{t}>{\;}$, we find 
$$a'{\;}+{\;}{{(e^{a}-1)}\over r}{\;}+{\;}{\dot a}{\;}e^{(a-b)/2}{\;}{\,} 
={\;}{\,}0{\quad}.\eqno(3.14)$$
\noindent
Using Eq.(3.8), one has
$$e^{(b-a)/2}{\;}{\,} 
={\;}{\,}-{\;}{{\dot m}\over m'}{\;}{\,} 
={\;}{\,}-{\;}(k_{r})_{\omega}{\;}(k_{t})_{\omega}{\;}e^{-a}{\quad},
\eqno(3.15)$$
\noindent
that is,
$$e^{b(t,r)}{\;}{\,}
={\;}{\,}\biggl({\dot m\over m'}\biggr)^{2}{\;}
\biggl(1-{2m\over r}\biggr)^{-1}{\quad}.\eqno(3.16)$$
\noindent
Then Eqs.(3.12,15) imply that
$$\theta_{\omega}{\;}{\,}={\;}{\,}\theta_{\omega}(m){\quad},\eqno(3.17)$$
\noindent
denoting an arbitrary function of $m{\,}$.  Finally, one arrives at 
the Vaidya solution [7,14], in the form quoted in Eq.(2.34,35),
$$ds^{2}{\;}{\,} 
={\;}-{\;}\biggl({\dot m \over m'}\biggr)^{2}{\;}
{\biggl(1-{{2m(t,r)}\over r}\biggr)}^{-1}{\;}dt^{2}{\;}
+{\;}\biggl(1-{{2m(t,r)}\over r}\biggr)^{-1}{\;}dr^{2}{\,}
+{\,}r^{2}{\;}d\Omega^{2}{\quad},\eqno(3.18)$$ 
\noindent
describing the background space-time $\gamma_{\mu\nu}$ which results
from the energy-momentum tensor of the high-frequency black-hole
radiation.\par
\smallskip
\indent
A change of variables: ${\,}(t,r)\rightarrow(u,r)$ can also be found
(see below), such that the line element [7] is of the
Eddington-Finkelstein type [9]:
$$ds^{2}{\;}{\,} 
={\;}{\,}-{\;}\biggl(1-{{2m(u)}\over r}\biggr){\,}du^{2}{\,}
-{\,}2{\,}du{\,}dr{\,}+{\;}r^{2}{\;}d\Omega^{2}{\quad}.\eqno(3.19)$$
\noindent
Then radially-outgoing null geodesics are precisely paths of constant
$u{\,}$.  The function $m$ is now independent of $r$ and constant along
outgoing null rays.  In the generic case that ${\,}(dm/du)$ is not
known, it has proved impossible to diagonalise the Vaidya metric and
to write $u$ as an explicit function of $t$ and $r{\,}$.  Since 
${\dot m}<0$ and $m'=(\partial m/\partial r)>0{\,}$, one finds that,
along lines $\{u ={\rm constant}\}$, $r$ increases with increasing 
$t{\,}$.  As in the fixed-mass Schwarzschild solution, the Vaidya
metric, in the form (3.18), has a coordinate singularity where 
${\,}r=2m(t,r)$.  But, from the $(u,r)$ form (3.19), one can see that 
the apparent singularity in the metric (3.18) at $r=2m(u)$ is only a
coordinate singularity [14].  Further [14], the surface $\{r=2m(u)\}$
is space-like, lying to the past of the region $\{r>2m(u)\}$, since
$(dm/du)>0{\,}$.  In fact, the geometry in the region $\{r<2m(u)\}$ 
(if such a region exists in the 'space-times' considered here, as generated
through solution of a boundary-value problem) would gradually deviate
from the Vaidya form, as one moves to the past by (say) reducing $u$
while holding $r$ fixed, since one would reach the region of
strong-field gravitational collapse.  This region can still be
described by the diagonal metric (2.34), but the full field equations
enforce a more complicated coupled solution.  Provided that the
complexified boundary-value problem, outlined in Sec.1, is well-posed for
a time-separation-at-infinity $T={\mid}T{\mid}\exp(-i\theta)$, for
$0<\theta\leq\pi/2{\,}$, then the full (complex) Einstein/scalar
classical solution studied here will be regular at the spatial origin
$r=0{\,}$.  Indeed, a solution would then be regular everywhere (with
respect to suitable coordinate charts) in the region between the
initial hypersurface $\Sigma_I$ and the final hypersurface 
$\Sigma_F{\,}$.  Since we are considering the case in which both 
$\Sigma_I$ and $\Sigma_F$ are diffeomorphic to ${\Bbb R}^3$, the 
solution should be regular on a region of the form 
$I\times{\Bbb R}^3$, where $I$ denotes the closed interval 
$[0,{\mid}T{\mid}]$.\par
\smallskip
\indent
Note further that the regularity of the boundary data 
$(h_{ij}{\,},\phi)$, as posed on $\Sigma_I$ and $\Sigma_F{\,}$, in the
spherically-symmetric case, implies that the boundary value $m_{I,F}(r)$
obeys $2{\,}m_{I,F}(r)<r{\,}$ for all $r>0{\,}$.  Equality only holds 
at the centre of symmetry $r=0{\,}$.\par
\smallskip
\indent
We now relate the Vaidya metric, as given in Eq.(3.18), to other
coordinate forms of the Vaidya geometry.  From Eq.(3.7) and from
differentiating Eq.(3.15) with respect to $r{\,}$, we find
$$\biggl({m''\over m'}{\;}-{\;}{{\dot m'}\over{\dot m}}\biggr){\;}
\biggl(1-{{2m}\over{r}}\biggr){\;}{\,} 
={\;}{\,}-{\;}{2m\over{r^{2}}}{\quad}.\eqno(3.20)$$
\noindent
This can be rearranged in the form
$${{\partial_{t}\Bigl(m'{\,}\Bigl(1-{{2m}\over{r}}\Bigr)\Bigr)}\over
{\partial_{r}\Bigl(m'{\,}\Bigl(1-{{2m}\over{r}}\Bigr)\Bigr)}}{\;}{\,} 
={\;}{\,}{{\dot m}\over{m'}}{\quad},\eqno(3.21)$$
\noindent
which has the solution
$$m'{\;}\biggl(1-{{2m}\over{r}}\biggr){\;}{\,}={\;}{\,}f(m){\quad},
\eqno(3.22)$$
\noindent
where $f(m)\geq 0$ is arbitrary [7].  Eq.(3.16) can now be rewritten
using Eq.(3.22), to give
$$e^{b(t,r)}{\;}{\,} 
={\;}{\,}e^{2\psi(t,r)}{\;}\biggl(1-{{2m(t,r)}\over{r}}\biggr){\quad},
\eqno(3.23)$$
\noindent
where ${\,}e^{2\psi(t,r)}{\,}$ is defined as
$$e^{2\psi(t,r)}{\;}{\,} 
={\;}{\,}\biggl({{\dot m}\over{f(m)}}\biggr)^{2}{\quad}.\eqno(3.24)$$ 
\noindent
Hence, the 4-metric can be written in the form 
$$ds^{2}{\;}{\,} 
={\;}{\,}-{\,}e^{2\psi(t,r)}{\;}\biggl(1-{{2m(t,r)}\over{r}}\biggr){\;}
dt^{2}{\;}
+{\;}\biggl(1-{{2m(t,r)}\over{r}}\biggr)^{-1}{\,}dr^{2}{\;}
+{\;}r^{2}{\;}d\Omega^{2}{\quad}.\eqno(3.25)$$
\noindent
The Vaidya model in the context of black-hole radiation has, for
example, been studied by Hiscock [15].  If one chooses $f(m)=-\dot m$,
then in Eq.(3.24) one has $\psi(t,r)=0$, whence the Vaidya metric of
Eq.(3.25) takes a particularly simple form. \par
\smallskip
\indent
Different choices of the function $f(m)$ correspond to different
physical models; once $f(m)$ is specified, one determines $m$ as a
function of $t$ and $r{\,}$.  One can consider the complexified 
boundary-value problem in the case (say) that exactly
spherically-symmetric initial data $(\gamma_{ij}{\,},\Phi)_I$ are
given on the initial hypersurface $\Sigma_I{\,}$, whereas on the final 
hypersurface $\Sigma_F{\,}$, the data consist of a background 
spherically-symmetric part $(\gamma_{ij}{\,},\Phi)_F$, together with 
weak 'linear-order' fluctuations $(h^{(1)}_{ij},\phi^{(1)})_F{\,}$, 
which correspond in the classical theory to emitted gravitons and 
massless scalar particles.  From experience with real elliptic partial 
differential equation theory, one might not unreasonably expect a 
unique classical solution to this Dirichlet boundary-value problem 
[16,17].  Hence, in particular, in the context of the high-frequency 
approximation of Sec.2, and of the Vaidya description of the 
corresponding outgoing-wave-source gravitational field of the present 
Section 3, one would expect that (for example) the detailed 
high-frequency coefficients 
$A_{\omega\ell m}(t,r,\Omega),{\,}(A_{\mu})_{\omega\ell mP}(t,r,\Omega)$ 
and $(A_{\mu\nu})_{\omega\ell mP}(t,r,\Omega)$ of Eqs.(2.25-27) should 
be determined by the above Dirichlet boundary data.  Similarly, the
'free' function $f(m)$ of Eq.(3.22) in the Vaidya description should
also be determined, and indeed $f(m)$ should be related to the
detailed quantities $A_{\omega\ell m}{\,}$, etc., above.\par
\smallskip
\indent
At late times and at correspondingly large radii $r{\,}$, the
semi-classical mass-loss formula should hold to great accuracy:
$$-{\;}{\dot m}{\;}{\,}={\;}{\,}{{\alpha(m)}\over{m^{2}}}{\quad},\eqno(3.26)$$
\noindent
where $\alpha(m)$ effectively accounts for the number of particles
light enough to be emitted by a hole of mass $m{\,}$, and $\alpha(m)$
increases with decreasing $m$ (here allowing for a more general model 
than our Einstein/massless-scalar case).  This follows since, as
described in Sec.3 of [5], the Bogoliubov coefficients for the quantum
evaporation of the black hole [17a] are given, for practical purposes,
by the standard expression
$${\mid}\beta_{s\omega\ell m}{\mid}^{2}\; =\; \Gamma_{s\omega\ell m} 
(\tilde m)\, \Bigl(e^{4\pi\tilde m} - (-1)^{2s}\Bigr)^{-1}. \eqno(3.27)$$
Here, $\Gamma_{s\omega\ell m}(\tilde m)$ is the transmission
probability over the centrifugal barrier of the black hole for a mode
with spin $s$, frequency $\omega$ and angular quantum numbers $(\ell,
m)$, and $\tilde m = 2M\omega$ is dimensionless, $M$ being the
space-like or total ADM (Arnowitt-Deser-Misner) mass of the space-time
[9].  The original derivation of Eq.(3.27) was in the context where
the black-hole singularity was taken to persist at late times.  But,
because of the very-high-frequency (adiabatic) method through which the
above expression for ${\mid}\beta_{s\omega \ell m}{\mid}^2$ was
calculated, it should still be valid (up to minute corrections) in the
case presently being studied, in which there is assumed to exist a
smooth final boundary $\Sigma_F$ with topology ${\Bbb R}^3$.  The
derivation of Eq.(3.26) then follows as usual.\par
\smallskip
\indent
In particular, consider the late-time behaviour appropriate to our
massless-field model, in which 
$\alpha(m)=\alpha_{0}={\rm constant}$.  The large-$r$ solution to 
Eq.(3.26), in the region where $m>0{\,}$, is of the form
$$m(u){\;}{\,} 
\simeq{\;}{\,}\Bigl((M_{I})^{3}{\,}+{\,}3{\,}\alpha_{0}{\,}
(u_{2}-u)\Bigr)^{1\over3}{\quad},\eqno(3.28)$$
\noindent
where $M_I$ and  $u_{2}$ are constants and $u{\,}\simeq{\,}(t-r)$ 
at large $r{\,}$.  We set ${\,}m(u_{2})=M_{I}$ for some fixed 
$u_{2}{\,}$, so that, as ${\,}u_{2}\rightarrow -\infty{\,}$, 
the space-like (ADM) and null (Bondi) masses [9] agree.\par
\smallskip
\indent 
Now introduce a null coordinate ${\,}u=u(t,r){\,}$, which agrees 
asymptotically with the above requirement 
${\,}u{\,}\simeq{\,}(t-r){\,}$, but which is defined everywhere, 
{\it via} the transformation 
$$du{\;}{\,}
={\;}{\,}\biggl(1{\,}-{\,}{{2m(t,r)}\over{r}}\biggr)^{-1}{\;}
\biggl(-{\,}{{\dot m}\over m'}\;dt-dr\biggr){\;}{\,}
={\;}{\,}-{\,}{{\dot m}\over{f(m)}}{\;}dt{\;}
-{\,}\biggl(1{\,}-{\,}{{2m(t,r)}\over{r}}\biggr)^{-1}{\;} 
dr{\quad}.\eqno(3.29)$$
\smallskip
\noindent
It may be verified that this definition is integrable, as follows:  
Using Eq.(3.22), one can re-write Eq.(3.29) in the form
$$du{\;}{\,}
={\;}-{\,}\biggl({{\dot m}\over{f(m)}}\biggr){\;}dt{\,}
-{\,}\biggl({{m^{\prime}}\over{f(m)}}\biggr){\;}dr{\;}{\,}
={\;}-{\,}\biggl({{dm}\over{f(m)}}\biggr){\;}{\,}
={\;}{\,}d\bigl({\mu}(m)\bigr){\quad},\eqno(3.30)$$
where we define
$${\mu}(m){\;}{\,}
={\;}{\,}-{\,}\int{{dm}\over{f(m)}}{\quad}.
\eqno(3.31)$$
It is then straightforward to apply the coordinate transformation
implicit in Eq.(3.30), to derive the 'null form' (3.19) of the Vaidya
metric from the alternative diagonal form (3.18).\par
\smallskip
\indent
We are now in a position to make further contact with the more
detailed treatment in Sec.2 of the high-frequency expansions
(2.25-27) for massless spin-0, spin-1, and spin-2 fields.  In the 
coordinate system $(u,r,\theta,\phi)$ of Eq.(3.19), we write out 
Eqs.(2.25-27) in the form
$$\phi^{(1)}(u,r,\Omega){\;}{\,} 
={\;}{\,}\sum_{\ell m}{\;}\int^{\infty}_{0}{\,}d\omega{\;}{\,} 
\Bigl[A_{\omega\ell m}(u,r,\Omega){\;}e^{i\theta_{\omega}(u,r)} 
+{\rm c.c.}\Bigr]{\quad},\eqno(3.32)$$
\noindent
etc.  The only non-zero component of the null vector
$(k^{\mu})_{\omega}$ is $(k^{r})_{\omega}$ [see Eq.(3.34) below], 
which, by Eq.(2.57), is in principle an arbitrary function of $u{\,}$.  
Thus, the radiation, corresponding to outgoing waves at null infinity, 
travels freely along $\{u={\rm constant}\}$ light cones.  Further, at
any point in the Vaidya space-time, a local observer finds only one 
direction in which the radiant energy is flowing.  Eq.(2.38) can now 
be solved to give
$$A_{s\omega\ell mP}(r,u,\Omega){\;}{\,} 
={\;}{\,}{{h_{s\omega\ell mP}(u,\Omega)}\over{r}}{\quad},\eqno(3.33)$$
\noindent
where ${\,}h_{s\omega\ell mP}(u,\Omega)$ is an arbitrary, dimensionless
complex function.  By this means, the coefficients 
${\,}A_{s\omega\ell mP}{\,}$ can be related to the distribution of 
weak-field massless-scalar, Maxwell and spin-2 graviton data on the 
final surface $\Sigma_F{\,}$.\par 
\smallskip
\indent
In the $(u,r,\theta,\phi)$ coordinate system, the only non-zero
component of the Ricci tensor is
$$R_{uu}{\;}{\,}
={\;}{\,}-{\;}{2\over{r^{2}}}{\;}{{dm}\over{du}}{\;}{\,} 
={\;}{\,}8\pi <T_{uu}>{\quad}.\eqno(3.34)$$
\noindent
Then Eq.(2.65) gives
$$\eqalign{-{\;}{1\over{4\pi{\,}r^{2}}}{\;}{{dm}\over{du}}{\;}{\,}&
={\;}{\,}2{\,}\sum_{s\ell mP}{\;}c_{s}{\;}\int^{\infty}_{0}{\;}
d\omega{\;}{\,}
\bigl[(k_{u})_{\omega}\bigr]^{2}{\;}{\mid}A_{s\omega\ell mP}{\mid}^{2}\cr
&={\;}{\,}{{2}\over{r^{2}}}{\;}\sum_{s\ell mP}{\;}c_{s}\;
\int^{\infty}_{0}{\;}d\omega{\;}{\,}\bigl[(k_{u})_{\omega}\bigr]^{2}{\;}
{\mid}h_{s\omega\ell mP}(u,\Omega){\mid}^{2}{\quad}.\cr}\eqno(3.35)$$
\noindent
Hence, $m(u)$ can only decrease as $u$ increases; the perturbation
amplitudes are non-zero if and only if the mass $m$ is changing.  Just
as the description in Eq.(3.33) of the coefficients 
$A_{s\omega\ell mP}$ in the high-frequency approximations (2.25-27) 
leads to a relation between $h_{s\omega\ell mP}$ and the perturbative 
spin-0, 1 and 2 data on the final surface $\Sigma_F{\,}$, so Eq.(3.35) 
gives $(dm/du)$ and hence $m(u)$ for the 'background part' of the
classical solution, in terms of the coefficients 
$A_{s\omega\ell mP}{\,}$.  That is, in setting up, as final data 
for gravitational collapse, the 'background part' 
$(\gamma_{ij}{\,},\Phi)_{\Sigma_F}{\,}$, together with the
perturbative part, one should choose the radial dependence of the 
late-time background part $\gamma_{ij}$ to allow for the mass 
function $m(u)$ corresponding to Eq.(3.33) for the given particle 
species and spins.\par
\smallskip
\indent 
Provided that the Lorentzian time-interval $T$ at infinity is
sufficiently large, one expects to study background 3-geometries
$\gamma_{ij}$ on the final hypersurface $\Sigma_F{\,}$, which are nearly
flat out to a certain large radius $r=R_{1}{\,}$, corresponding to the
edge of the region in which the radiation reaches $\Sigma_F{\,}$.  
For ${\,}r>R_{1}{\,}$, one expects $\gamma_{ij}$ to correspond to a 
slowly-varying Vaidya metric, with ${\,}m(u)=m(T-r){\,}$ gradually 
increasing out to a radius $R_{2}$ which corresponds roughly to the 
beginning of the radiation.  At radii $r>R_{2}$ on the final surface 
$\Sigma_F{\,}$, the function $m(u)$ should be approximately equal 
to $M_{I}{\,}$, the conserved ADM mass of the system.\par
\smallskip
\indent
At the high-energy end of the emission spectrum, when the black hole
approaches the Planck scale, the (thermal) mass-loss rate as given by
Eq.(3.26) breaks down.  The amount of energy emitted by the black hole
in the final stages of the evaporation will be comparable to its mass,
$\omega \sim m$.  To account for the small-mass behaviour of the black
hole, therefore, the micro-canonical decay rate must be considered.
The micro-canonical approach is generally more desirable, as the
thermal equilibrium between a black hole and the exterior radiation is
unstable, due to a negative specific heat in the canonical ensemble
[18].  In addition, there is no information loss in the black-hole
evaporation in the micro-canonical picture, as energy is
conserved. \par
\smallskip
\indent
For the low-frequency quanta ($\omega\ll M$) characteristic of the
majority of the evaporation process, however, the canonical and
micro-canonical ensembles are almost equivalent, and one obtains a
Planck-like number spectrum and the decay rate Eq.(3.26).  The
micro-canonical decay rate for small $m$ has the form [19] 
$$\eqalign{-\dot {m} \,& =3\,f(m)\cr
& \sim \Bigl({{\lambda  m_{pl}}\over {t_{pl}}}\Bigr)\cr
&= \Bigl({{m}\over{m_{pl}}}\Bigr)^{6}\; \exp\Bigl(-4\pi
{{m^{2}}\over{m_{pl}^{2}}}\Bigr),\cr} \eqno(3.36)$$
where $\lambda$ is a numerical constant and $m_{pl},t_{pl}$ denote the
Planck mass and Planck time, respectively.  This further equation does
not have the bad behaviour as $m\rightarrow 0$, associated with
Eq.(3.26).  Indeed, the free function $f(m)$ will naturally have the
corresponding form for small $m$, as follows from a dimensional
analysis of the field amplitudes.  \par
\medskip
\noindent
{\bf 4. Conclusion}
\medskip
\indent
We have seen in Sec.2 how the averaged effective energy-momentum
tensor $<T^{\mu\nu}>_{EFF}$ is calculated, describing, over scales of
several radiation wavelengths, the way in which wave-like fluctuations
in the spin-0 (scalar) field $\phi^{(1)}$ and the spin-2 (graviton)
part of the linearised gravitational field $h^{(1)}_{\mu\nu}$
contribute quadratically as sources for the 'background'
spherically-symmetric 4-metric and scalar field 
$(\gamma_{\mu\nu}{\,},\Phi)$.  A similar description holds for the 
spin-1 Maxwell field.  While this contribution is small at any one 
time, it persists with a comparable magnitude during the whole time 
$t_{0}$ during which the black hole radiates.  Thus, its effects, 
particularly on the spherically-symmetric background metric 
$\gamma_{\mu\nu}$, accrue secularly; indeed, the averaged contribution 
$<T^{\mu\nu} >_{EFF}$ is precisely such as to determine the rate of 
loss of mass $(-\dot m)$ in the familiar fashion, leading to the 
eventual disappearance of the central concentration of mass, when one 
works with a complexified time-interval
$T={\mid}T{\mid}\exp(-i\theta)$, 
with ${\,}0<\theta\leq\pi/2{\,}$, for which one expects a classical
solution which is regular between the initial hypersurface $\Sigma_I$
and final hypersurface $\Sigma_F{\,}$.\par
\smallskip
\indent
Such an averaged effective energy-momentum source leads to an
approximate space-time geometry $g_{\mu\nu}$ of the Vaidya type, as
described in Sec.3, valid in the space-time region containing the
outgoing radiation.  This Vaidya description is in turn essential in
the treatment of adiabatic radial mode equations, as in [3-5].
Subsequently, in [20], we have generalised the boundary-value 
treatment in [1-4], which refers to quantum amplitudes with only 
spin-0 perturbative data on the final hypersurface $\Sigma_F{\,}$,
at a late time $T{\,}$.  In [20], we treat the other bosonic cases of 
spin-1 and spin-2 final data, by means of similar but more 
complicated methods.  The fermionic massless 
spin-${1}\over{2}$ case is treated in [21], and a treatment of the 
remaining fermionic spin-${3}\over{2}$ case is in preparation [22]; 
this is needed as part of the treatment of locally-supersymmetric 
models.  In all these examples, understanding of the Vaidya
description is essential.\par
\parindent = 1 pt
\medskip
{\bf Acknowledgments}
\medskip
We are grateful to a referee for constructive suggestions.\par
\medskip
{\bf References}
\medskip
\indent
[1] A.N.St.J.Farley and P.D.D'Eath, 
Phys Lett. B {\bf 601}, 184 (2004).\par          
\indent
[2] A.N.St.J.Farley and P.D.D'Eath, 
Phys Lett. B {\bf 613}, 181 (2005).\par          
\indent 
[3] A.N.St.J.Farley and P.D.D'Eath, 
`Quantum Amplitudes in Black-Hole Evaporation: I. Complex Approach', 
submitted for publication (2005).\par
\indent
[4] A.N.St.J.Farley and P.D.D'Eath, 
`Quantum Amplitudes in Black-Hole Evaporation: II. Spin-0 Amplitude', 
submitted for publication (2005).\par
\indent      
[5] A.N.St.J.Farley and P.D.D'Eath, 
'Bogoliubov Transformations in Black-Hole Evaporation', 
submitted for publication (2005).\par
\indent
[6] H.Stephani et al., 
{\it Exact Solutions to Einstein's Field Equations}, 2nd. ed, 
(Cambridge University Press, Cambridge) (2003).\par
\indent
[7] P.C.Vaidya, Proc. Indian Acad. Sci. {\bf A33}, 264 (1951).\par 
\indent
[8] R.Isaacson, Phys. Rev. {\bf 166}, 1263, 1272 (1968).\par
\indent
[9] C.W.Misner, K.S.Thorne and J.A.Wheeler, {\it Gravitation},
(Freeman, San Francisco) (1973).\par
\indent
[10] J.Wess and J.Bagger, {\it Supersymmetry and Supergravity},
2nd. edition, (Princeton University Press, Princeton) (1992).\par
\indent
[11] C.M.Bender and S.A.Orszag, 
{\it Advanced Mathematical Methods for Scientists and Engineers} 
(Springer, New York) (1999).\par
\indent
[12] T.Regge and J.A.Wheeler, Phys. Rev. {\bf 108}, 1063 (1957).\par 
\indent
[13] P.D.D'Eath, {\it Supersymmetric Quantum Cosmology}, 
(Cambridge University Press, Cambridge) (1996).\par
\indent
[14] R.W.Lindquist, R.A.Schwartz and C.W.Misner, 
Phys. Rev. {\bf 137}, 1364 (1965).\par
\indent
[15] W.A.Hiscock, Phys. Rev D {\bf 23}, 2813, 2823 (1981).\par
\indent
[16] P.R.Garabedian, {\it Partial Differential Equations}, 
(Wiley, New York) (1964).\par
\indent
[17] W.McLean, 
{\it Strongly Elliptic Systems and Boundary Integral Equations}, 
(Cambridge University Press, Cambridge) (2000);
O.Reula, 'A configuration space for quantum gravity and solutions to
the Euclidean Einstein equations in a slab region', 
Max-Planck-Institut f\"ur Astrophysik, {\bf MPA} 275 (1987).\par
\indent
[17a] S.W.Hawking, Commun. Math. Phys. {\bf 43}, 199 (1975);
N.D.Birrell, P.C.W.Davies, {\it Quantum Fields in Curved Space},
(Cambridge University Press, Cambridge) (1982); V.P.Frolov,
I.D.Novikov, {\it Black Hole Physics}, (Kluwer Academic, Dordrecht)
(1998).\par 
\indent
[18] S.W.Hawking, Phys. Rev. D {\bf 13}, 191 (1976).\par
\indent
[19] R.Casadio, B.Harms, and Y.Leblanc, Phys. Rev. D {\bf 58}, 044014
(1998).\par
\indent
[20] A.N.St.J.Farley and P.D.D'Eath, 
Class. Quantum Grav. {\bf 22}, 2765 (2005).\par 
\indent
[21] A.N.St.J.Farley and P.D.D'Eath, 
Class. Quantum Grav. {\bf 22}, 3001 (2005).\par 
\indent
[22] A.N.St.J.Farley and P.D.D'Eath, 
'Spin-3/2 Amplitudes in Black-Hole Evaporation', in progress.\par

\end